\documentclass[journal=nalefd,manuscript=letter]{achemso}
\usepackage{chemformula} 
\usepackage[T1]{fontenc} 
\usepackage[colorlinks,linkcolor=red]{hyperref}

\author{Jun-Shan Si}
\author{Hongxing Li}
\author{Bin-Guang He}
\author{Zi-Peng Cheng}
\author{Wei-Bing Zhang}
\email{zhangwb@csust.edu.cn}
\affiliation{Hunan Provincial Key Laboratory of Flexible Electronic Materials Genome Engineering, School of Physics and Electronic Sciences, Changsha University of Science and Technology, Changsha 410114, People's Republic of China.}

\title {Revealing the Underlying Mechanisms of Stacking Order and Interlayer Magnetism in Bilayer CrBr$_3$}


\begin{document}

\begin{tocentry}
\center
\includegraphics{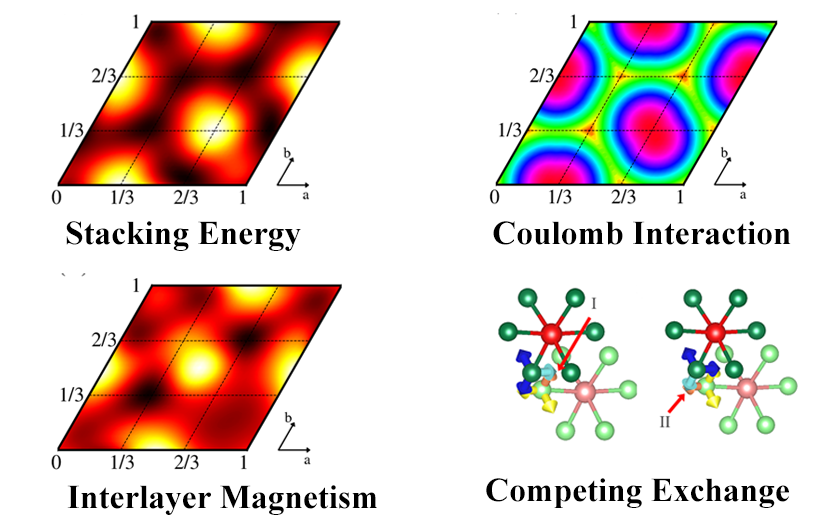}

\end{tocentry}

\begin{abstract}
Aiming to clarify the mechanisms governing the interlayer magnetic coupling, we have investigated the stacking energy and interlayer magnetism of bilayer CrBr$_3$ systemically. The magnetic ground states of bilayer CrBr$_3$ with different R-type and H-type stacking orders are established, which is found to be in good agreement with recent experiment (\emph{Science} $\mathbf{366}$,983(2019)).Further analyses indicate that the stacking energy is mainly determined by the Coulomb interaction between the interlayer nearest-neighbor Br-Br atoms. While interlayer magnetism can be understood by a competition between super-super-exchange interactions involving $t_{2g}$-$t_{2g}$ and $t_{2g}$-$e_g$ orbitals and semi-covalent exchange interactions of $e_g$-$e_g$ orbitals. Our studies give an insightful understanding for stacking order and interlayer magnetism of bilayer CrBr$_3$, which should be useful to understand quantum confinement effect of other layered magnets in two-dimensional limit.
\end{abstract}

The discovery of intrinsic ferromagnetism (FM) in atomically thin crystals has triggered increasing interest in two-dimensional (2D) magnetism \cite{exp_crI3,bilayer_fe,mag_nature,NN_mag_heter,zhang_sci,xu_2020}. As the first Ising-type single-layer ferromagnetic semiconductor realized experimentally \cite{exp_crI3}, CrI$_3$ has attracted special interest due to the novel quantum confinement effect. Bulk CrI$_3$ shows ferromagnetic (FM) order both within and between layers \cite{cm,cryst,zhang_CrI3}, whereas its few-layer samples possess interlayer antiferromagnetism. \cite{Li_NM,science_prob,song_nm} This layer-dependent interlayer magnetic order leads to a number of emerging exotic phenomena \cite{Klein1218_TMR,Song1214_GTM,nc,Jiang_nm,Huang_nn,jiang_NN,nature_second} including  giant  magnetoresistance in magnetic tunnel device and nonreciprocal second-order nonlinear optical effect. Although recent experimental \cite{Li_NM,science_prob,song_nm} and theoretical studies \cite{theory_nl,theory_prb,theory_prm,theory_ssc} have confirmed a clear correlation between the magnetic ground state and the stacking order, the underlying mechanism is not yet understood.\cite{fudan_2019}

CrBr$_3$ is another important vdW magnetic material, which has been known as the first ferromagnetic semiconductor.\cite{crBr3} The rise of 2D magnetism also intrigues extensive research on the magnetism of CrBr$_3$ in two-dimensional limit\cite{CrBr3_ne,fudan_2019,pnas}. The interlayer coupling in atomically thin CrBr$_3$ is found to be ferromagnetic \cite{CrBr3_ne,fudan_2019,pnas}, which is distinct from CrI$_3$. Significant difference between CrBr$_3$ and CrI$_3$ evokes the further understanding of interlayer magnetism. To reveal the underlying mechanism, Chen \emph{et al.} \cite{fudan_2019} have grown the bilayer CrBr$_3$ with various stacking orders through molecular beam epitaxy methods, and observed stacking-dependent interlayer magnetism directly. This interesting experimental result provides an excellent platform to study and verify the magnetic mechanism in van der Waals bilayers.

\begin{figure*}
\centering
\includegraphics[width=0.85\textwidth]{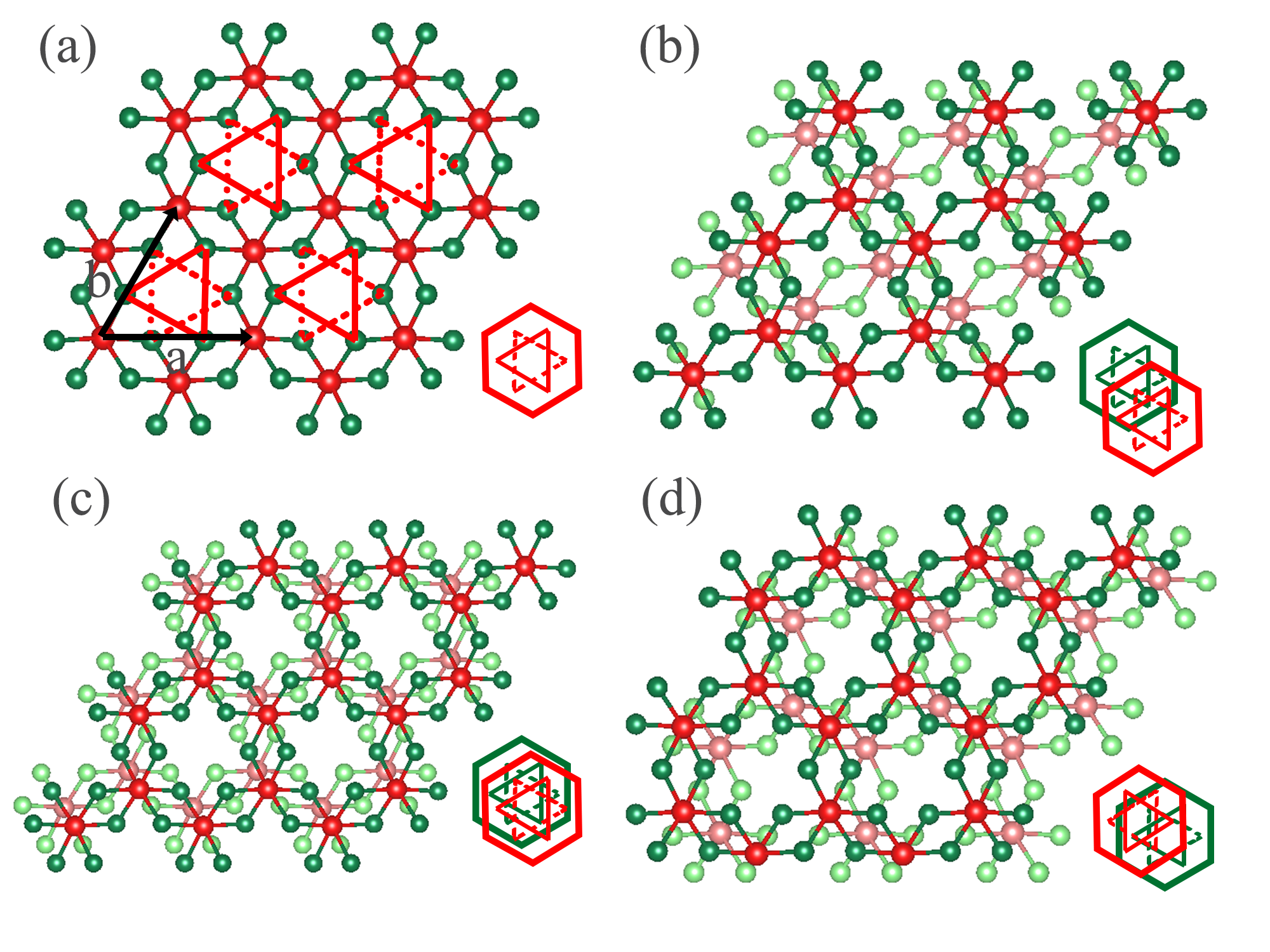}
\caption{\label{fig:stru}(Color online) The crystal structure of bilayer CrBr$_3$ with different stacking orders. (a) The top view of monolayer CrBr$_3$. Within the Cr honeycomb lattice, the top and bottom surfaces of Br atoms form single triangles but with opposite orientation, marked in solid and dotted red lines, respectively. The top view of three bilayer CrBr$_3$ with different stacking R$_{33}$, R$_{15}$, H$_{41}$ are shown in (b),(c) and (d) respectively. }
\end{figure*}

Similar with CrI$_3$, the bulk CrBr$_3$ has a low-temperature phase with the space group R$\bar{3}$ and a high-temperature phase with space group \emph{C2/m}.\cite{PhysRevB.60.16170,cm} As shown in Fig.~\ref{fig:stru}, Cr atoms of monolayer CrBr$_3$ are arranged in a honeycomb lattice and each atom is surrounded by an octahedron of six Br atoms. Within a single honeycomb formed by six Cr atoms, there are three top and bottom Br atoms marked by solid and dotted red triangles with opposite orientations, respectively. Due to the weak interlayer vdW interaction, different stacking orders are easy to form. As labeled in recent experimental work \cite{fudan_2019}, there are two kinds of stacking orders, namely R-type and H-type. Both layers align to the same orientation in the R-type but rotate by a 180$^{\circ}$ in H-type stacking order. To obtain the details of stacking energy and interlayer exchange energy, a uniform 6$\times$6 displacement vector grid in surface unit cell, which corresponds to 36 different stacking structures R$_{mn}$(H$_{mn}$), was considered in the calculation. R$_{mn}$(H$_{mn}$) means that the upper layer is shifted by $\frac{m}{6}\vec{a}$+$\frac{n}{6}\vec{b}$ relative to the AA-stacked R-type(H-type) bilayer, in which $\vec{a}$ and $\vec{b}$ are the surface unit vectors respectively. Three typical structures R$_{33}$, R$_{15}$ and H$_{41}$, which are prepared in recent experiment successfully, are shown in Fig.~\ref{fig:stru}-(b) to -(d) for clear illustration.

All DFT calculations are performed using the Vienna simulation package (VASP) code\cite{vasp_CMS,vasp_PRB} within the projector augmented-wave (PAW) method \cite{vasp_PAW,vasppawprb}. General gradient approximations (GGA) both in PBEsol \cite{pbesol} and PBE implementations \cite{PBE} are adopted as the exchange correlation functional. A plane-wave basis set with a cutoff energy of 700 eV and a $\Gamma$-centered 17$\times$17$\times$1 Monkhorst-Pack grid are used in the calculations. A simple rotationally invariant DFT + U method \cite{ldau} with effective on-site Coulomb interaction $U-J$ of 2.8 eV (U=3.9 eV, J=1.1 eV) is chosen for the Cr atoms to account for strong electronic correlations.\cite{theory_nl} Atomic positions are optimized with a force convergence tolerance of 3 meV/\AA. A vacuum spacing of 20\AA~ is used in the supercell to avoid interaction between images. The phonon calculations have been performed using the finite-displacement method, as implemented in the Phonopy code \cite{phonopy}, in which a 2$\times$2$\times$1 supercell is used.

\begin{figure*}
\centering
\includegraphics[width=0.85\textwidth]{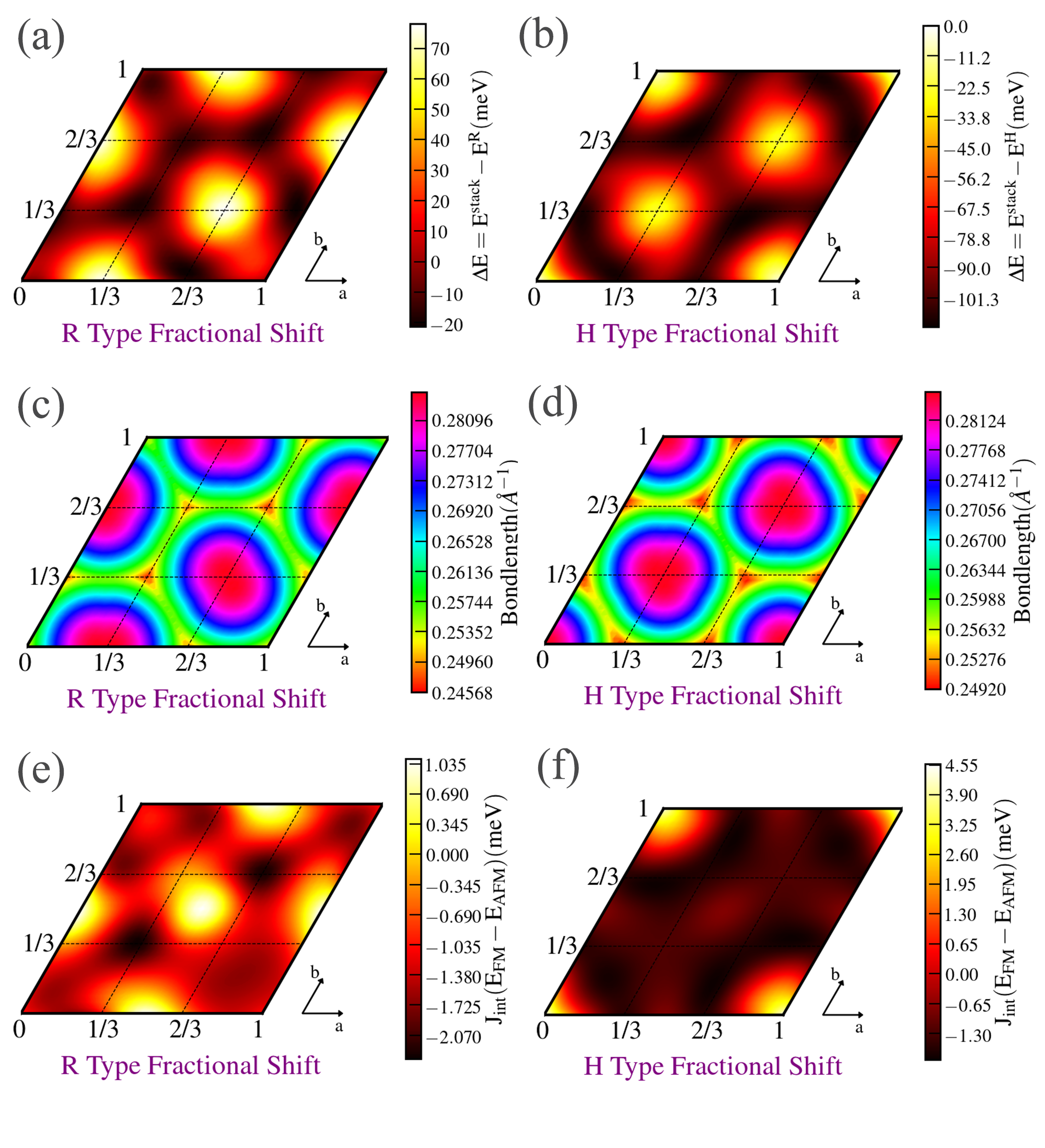}
\caption{\label{fig:pes}(Color online) The stacking energy, interlayer nearest-neighbor(NN) Br-Br distance and  interlayer exchange energy as a function of lateral shift, with respect to AA-stacking R-type and H-type bilayer CrBr$_3$. (a),(c) and (e) represent the results of R-type stacking configures, while the corresponding results for H-type are shown in (b),(d) and (f).The heat-maps were drawn by interpolating over the neighboring data points based on computational data obtained using a 6$\times$6 grid.  }
\end{figure*}

Starting from the fully relaxed R$_{22}$ structure, we constructed bilayer CrBr$_3$ with various stacking orders by rigidly shifting the upper atomic layer along $\vec{a}$ and $\vec{b}$ direction. The stacking energies, defined as the energy difference between different stacking and AA-stacking structures, are shown in Fig.~\ref{fig:pes}. We can find that the change trend of stacking energy for R-type CrBr$_3$ is similar to that of CrI$_3$,\cite{theory_nl} but the energy difference is much smaller. This implies that formation of stacking order in CrBr$_3$ is much easier, which also consists with the diversity of interlayer stacking orders found in recent MBE experiment.\cite{fudan_2019}

Moreover, we can also find that R$_{22}$ and R$_{24}$ locate at energy minimum of potential energy surface, which corresponds to the stacking order of R$\bar{3}$ and \emph{C2/m} bulk phases, respectively. It should be noticed that three structures R$_{33}$, R$_{15}$ and H$_{41}$ identified in recent experiments do not locate in the local minimum. However, the energy differences between various stacking orders are tiny, which implies a similar energetic stability. To further evaluate their stability, we also calculated the phonon spectrum of three kinds of structure. As shown in Fig.~\ref{fig:phonon}, there are essentially no imaginary frequencies in the whole Brillouin zone , which indicates that these stacking structures are dynamically stable and should be feasible in experiment.

To reveal the underlying mechanism governing stacking energy, we plot the variance of the reciprocal of the distance ($\frac{1}{r}$) between two nearest-neighbor(NN) Br atoms in different single-layers. As shown in Fig.~\ref{fig:pes}-(c) and (d), the variance of $\frac{1}{r}$ are highly similar to the stacking energy surfaces. Moreover, we can even judge the relative stacking energy using the distance of interlayer NN Br atoms. For example, a vary small barrier between R$_{22}$ and R$_{24}$ can be found in Fig.~\ref{fig:pes}-(c), which agrees with the stacking energy surface (Fig.~\ref{fig:pes}-(a)). Since the Coulomb potential of two atoms is known to be proportional to $\frac{1}{r}$, we thus conclude that the stacking energy is mainly contributed by the Coulomb interaction between the interlayer NN Br atoms.

We now turn our attention to the interlayer magnetic order of bilayer CrBr$_3$ with different stacking orders. The interlayer exchange energies, defined as the energy difference between the interlayer ferromagnetic and anti-ferromagnetic spin configurations, are calculated and shown in Fig.~\ref{fig:pes}-(e) and (f).  Since interlayer exchange energy is very subtle, we also used different methods to validate our results. Firstly, we check the variance of interlayer exchange energy with Coulomb interaction U in LDA+U methods. As shown in the Fig.~\ref{fig:plusu}, although the exchange energy change slightly but the interlayer magnetic order keeps invariant. Moreover, further calculations including the spin-orbital coupling (SOC) effect and using PBE functional have also been performed for comparison. As shown in Table.~\ref{tab:methods}, all methods predict the consistent interlayer magnetic order, indicating the reliability of the present investigation. Our calculations indicate both R$_{22}$ and R$_{24}$ structures, which correspond to the stacking order of bulk R$\bar{3}$ and \emph{C2/m} phases, show interlayer ferromagnetism.  This agrees with recent tunneling measurements,\cite{CrBr3_ne} in which the interlayer coupling in atomically thin CrBr$_3$ is found to be ferromagnetic. Moreover, R$_{33}$ and H$_{41}$ stacking was found to be antiferromagnetic and ferromagnetic, which is also consistent with the recent experimental results.\cite{fudan_2019} However, the experimentally identified anti-ferromagnetic R$_{15}$ phase was predicted to be interlayer ferromagnetic. Considered the agreement between experiment and theory about other stacking orders, further experimental verification about interlayer magnetism of R$_{15}$-stacking is highly required.

\begin{table*}
\caption{\label{tab:methods}}
Interlayer exchange energy in unit of meV of bilayer CrBr$_3$ with different stacking orders calculated using four methods.

\begin{tabular}{cccccccccccc}
 \hline
Type&PBE+U&	PBE+U+SOC&	PBEsol+U	&PBEsol+U+SOC\\
\hline
Hollow II (R$_{22}$)	&-1.896&	-2.146	&-2.305&	-2.001\\
Hollow III (R$_{24}$)&	-0.189&	-0.192	&-0.261&	-0.209\\
Special (R$_{33}$)	&0.942&	0.832&	1.077	&1.006\\
Bridge I (R$_{15}$)	&-1.076	&-1.242	&-1.312	&-1.103\\
Bridge I (H$_{41}$)	&-1.476	&-1.651	&-1.760&	-1.643\\
\hline
\end{tabular}
\end{table*}

\begin{figure*}
\centering
\includegraphics[width=0.85\textwidth]{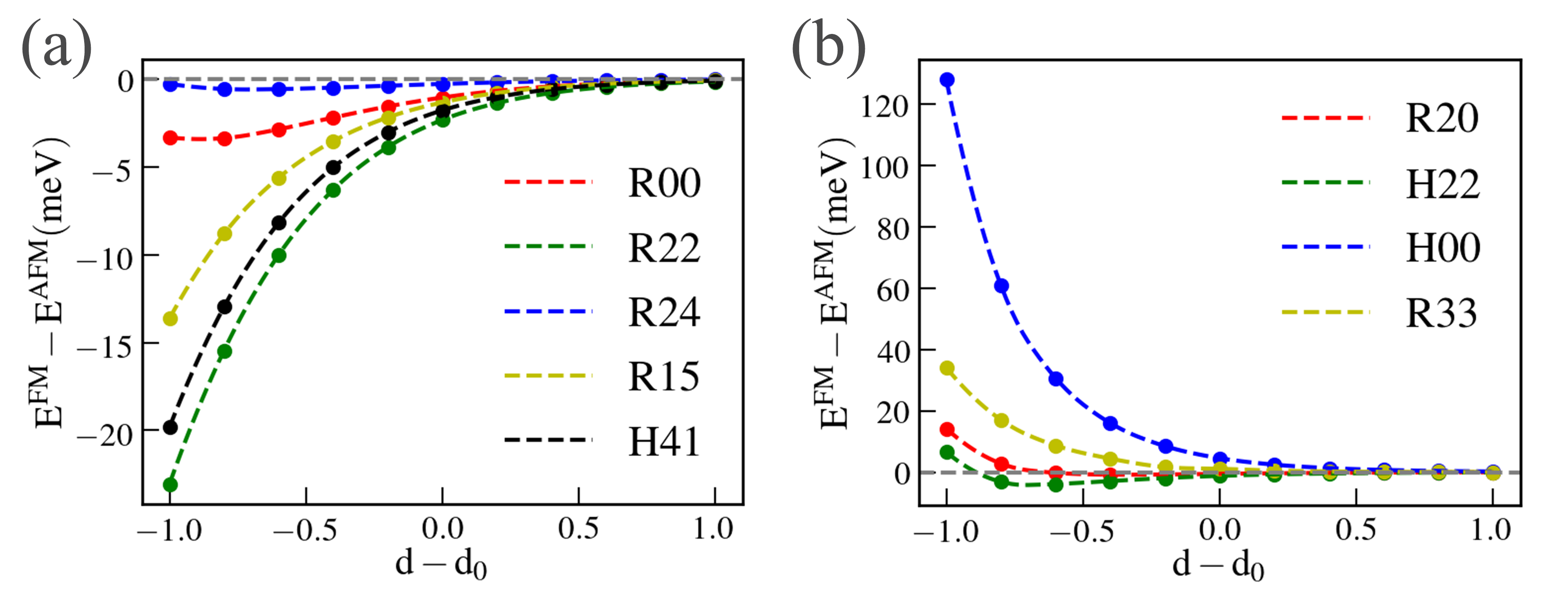}
\caption{\label{fig:distance}(Color online) The evolution of interlayer exchange energy with layer distances for different stacking orders. }
\end{figure*}

In addition, we also studied the influence of layer distance on interlayer magnetism. As shown in the fig.~\ref{fig:distance}-(a) and -(b). We can find a interesting stacking-dependent behaviour. For example,  the reduction of distance enhances interlayer ferromagnetism for R$_{22}$ and H$_{41}$, but enhances anti-ferromagnetism in case of R$_{33}$ and H$_{00}$ phase. Moreover, the change of layer distance will tune the interlayer magnetic state between FM and AFM of H$_{22}$ and R$_{20}$ stacking as shown in Fig.~\ref{fig:distance}-(b). This change trend is similar to the potential energy curve of diatomic molecules, which indicates that there may be two kind of competition \emph{i.e.} ferromagnetism and anti-ferromagnetism governing the interlayer magnetic coupling in bilayer CrBr$_3$.

\begin{figure*}
\centering
\includegraphics[width=0.85\textwidth]{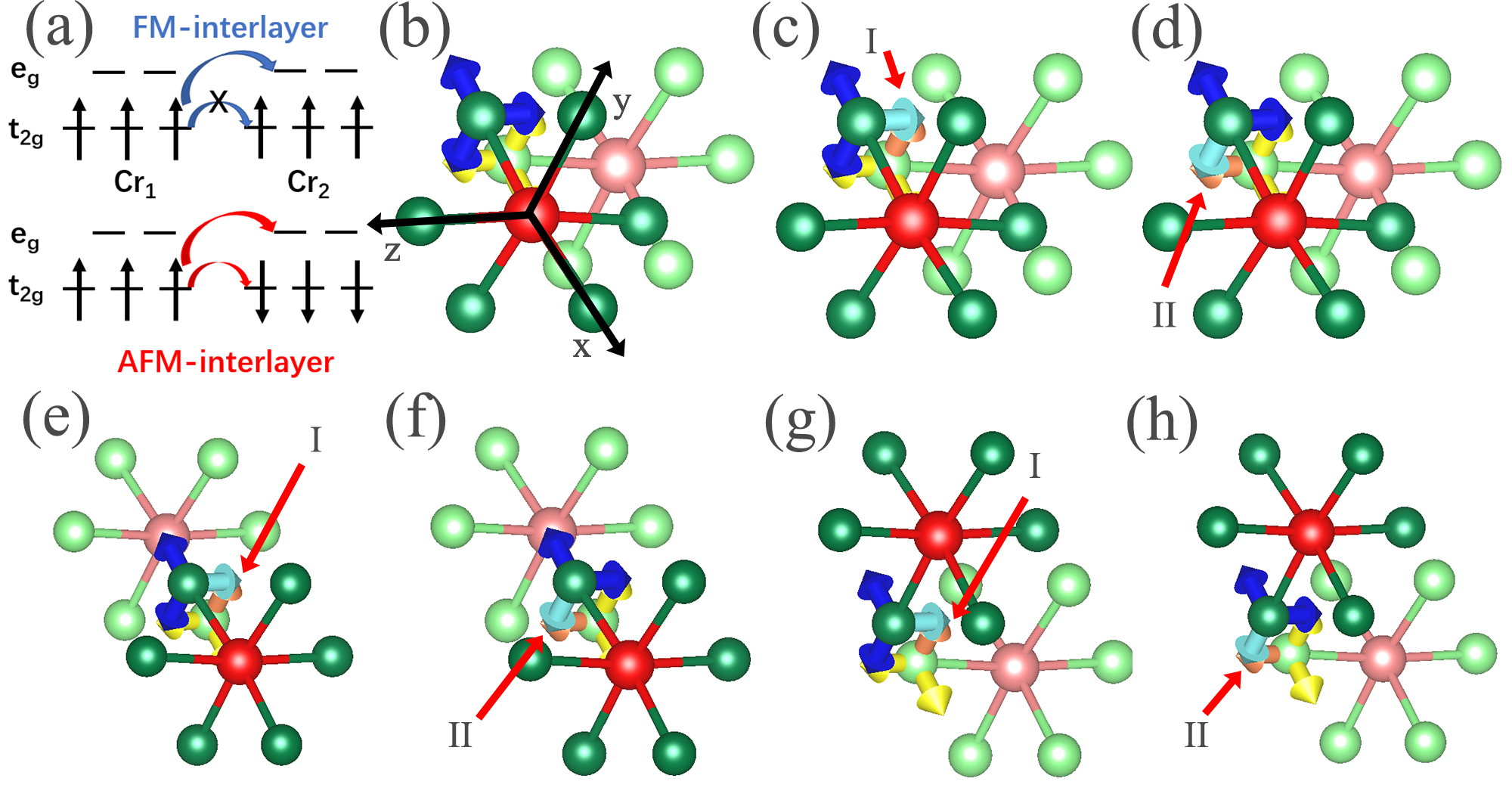}
\caption{\label{fig:sse}(Color online) The interlayer magnetic exchange mechanism. (a) A schematic diagram of the orbital dependent interlayer super-super-exchange interactions. Hopping of the form t$_{2g}$-t$_{2g}$ is prohibited in FM exchange (blue) but allowed in AFM exchange (red). Three directions of p orbital of the nearest-neighbors Br atoms are shown in (b). The nearest-neighbors interlayer interactions are shown in (c) and (d). (e)-(h) represent the next-nearest-neighbors interlayer interaction.}
\end{figure*}

Now, let's focus on the microscopic mechanism of interlayer magnetic coupling. As established in the earlier study, the Cr$^{3+}$ ions in CrBr$_3$ exhibit a high-spin t$_{2g}^3$e$_g^0$ electronic configuration.\cite{zhang_CrI3} The intra-layer magnetic coupling is predominated by the Cr-Br-Cr super-exchange interaction. For the interlayer magnetic coupling, there are several possible exchange interactions. Since the distance between Cr atoms in different layers are much larger than common super-exchange path, the cations are separated by two anions in the possible super-exchange path, forming an Cr–Br–Br–Cr interaction. The well-known Goodenough-Kanamori-Anderson(GKA) rule should be qualified to understand this interaction.\cite{goodenough}  Fig.~\ref{fig:sse}-(a) shows a schematic diagram of different exchange interactions between interlayer Cr atoms. t$_{2g}$-t$_{2g}$ transitions are prohibited for ferromagnetic coupling but allowed for anti-ferromagnetic coupling. On the other hand, due to the local Hund rule, the hopping of t$_{2g}$-e$_{g}$ leads to the ferromagnetic exchange coupling. In addition, although the two empty e$_{g}$-e$_{g}$ orbitals can't give super-exchange-like interaction, they can give anti-ferromagnetic $\sigma$-bond spin-spin interactions via a purely semi-covalent-exchange interaction.\cite{goodenough} Recent orbital-resolved magnetic exchange interaction calculations of CrI$_3$\cite{theory_prm} also give the similar pictures, in which t$_{2g}$-t$_{2g}$, t$_{2g}$-e$_{g}$, and e$_{g}$-e$_{g}$ interactions are found to be anti-ferromagnetic, ferromagnetic and anti-ferromagnetic, respectively.

Clearly, the interlayer magnetic coupling of CrBr$_3$ can be understood by the competition of these three interactions. According to the symmetry matching principle of molecular orbital, e$_g$ and t$_{2g}$ orbital of Cr-3d electron can hybridize with $\sigma$ and $\pi$ state of the ligands Br atoms, respectively. Since the $\sigma$ bond is known to be stronger than $\pi$ bond, the strength of e$_g$-e$_g$ interaction is expected to be greater than that of the t$_{2g}$-e$_g$, while the t$_{2g}$-t$_{2g}$ involves two p-$\pi$ orbitals should be weakest under the same conditions. This conclusion is also supported by recent research about interlayer magnetism of CrI$_3$.\cite{theory_prm}

Due to the intermediation role of Br-Br atoms, whether two p orbitals of interlayer NN Br atoms interact with each other is curial to form the possible exchange interaction. As shown above, the local coordinate system of octahedral environment is different from the global cartesian coordinate. The coordinate system of the two monolayers should be identical for R-type stacking but different for H-type stacking order, as shown in Fig.~\ref{fig:orb}. When the directions of two p orbitals intersect, the wavefunction of two p orbitals should be overlapped and possible hopping can happen (Fig.~\ref{fig:orb}-(b) and (c). However, if the directions of two p orbitals do not intersect, no interaction would occur due to their large distance.(Fig.~\ref{fig:orb}-(a) and (d)).

Taking the R$_{33}$ structure as an example, we illustrate the above idea to understand the interlayer magnetic coupling of bilayer CrBr$_3$. As shown in Fig.~\ref{fig:sse}-(b), there are three different directions for p- orbital of each Br atoms for nearest-neighbor(NN) Cr-Br-Br-Cr path, in which the blue and yellow arrows represent the p-orbital in upper and lower atomic layer, respectively. According to the above assumption, two groups of orbitals labeled by I (Fig.~\ref{fig:sse}-(c)) and II (Fig.~\ref{fig:sse}-(d)) are expected to interact with each other. And the corresponding d orbital, which hybridizes with the above p orbital, can be determined according to the principle of symmetry matching. For group I, both the p orbital of the upper and lower Br atom are perpendicular to the Br-I bond. These p-$\pi$-like molecular orbital will hybrid with the t$_{2g}$ orbital of Cr atom, which forms a t$_{2g}$-t$_{2g}$ interaction. For group II, the p-$\pi$-like orbital of the upper Br atom can only bind with the t$_{2g}$ orbital, while the p-$\sigma$-like orbital of the lower Br atom would form molecular orbital with the e$_g$ orbital of Cr atom. The interaction between the next-nearest-neighbors (NNN) Cr atoms is similar to that of NN. As shown in the Fig.~\ref{fig:sse}-(e) to Fig.~\ref{fig:sse}-(h), there are two non-equivalent Cr-Br-Br-Cr channels between the NNN Cr atoms. For channel 1 (Fig.~\ref{fig:sse}-(e) and -(f)), according to the previous analysis, it can be obtained that both groups I and II have t$_{2g}$-t$_{2g}$ interaction. For channel 2, group I and group II have t$_{2g}$-t$_{2g}$ and e$_g$-e$_g$ interaction, respectively. Totally, there were 6 t$_{2g}$-t$_{2g}$ and 6 t$_{2g}$-e$_g$ interactions between the NN Cr atoms in a single cell with R$_{33}$ structure containing 4 Cr atoms. And 14 t$_{2g}$-t$_{2g}$ and 6 e$_g$-e$_g$ interactions are present between the NNN Cr atoms. Since the t$_{2g}$-t$_{2g}$ interaction is much weaker, t$_{2g}$-e$_g$ and e$_g$-e$_g$ will be determinative. Considered the e$_g$-e$_g$ is stronger than t$_{2g}$-e$_g$ interaction and the number of both interactions are identical, the interlayer magnetic coupling of R$_{33}$ should be anti-ferromagnetic.

\begin{table*}
\caption{\label{tab:path}}
The interlayer NN and NNN exchange interactions between Cr atoms and the corresponding number for different stacking orders.

\begin{tabular}{ccccccccccccccccccc}
 \hline
Stacking&	States&	Path&	Number&	Type&	Number&	Type&	Number\\
\hline
R$_{22}$	&FM	&NN&	1&	t$_{2g}$-t$_{2g}$&	6&&\\		
&&NNN	&15	&t$_{2g}$-t$_{2g}$&	15	&t$_{2g}$-e$_{g}$&	30\\

R$_{24}$	&FM &	NN&	6&	t$_{2g}$-t$_{2g}$&	12&	t$_{2g}$-e$_{g}$&	24\\
&&		NNN&	7&	t$_{2g}$-t$_{2g}$	&7	&e$_{g}$-e$_{g}$	&7\\

R$_{33}$	&AFM	&NN&	3&	t$_{2g}$-t$_{2g}$&	6&	t$_{2g}$-e$_{g}$&	6\\
	&&	NNN&	7	&t$_{2g}$-t$_{2g}$&	14&	e$_{g}$-e$_{g}$&	6\\

H$_{41}$	&FM&	NN&	4&	t$_{2g}$-t$_{2g}$ &	8	&t$_{2g}$-e$_{g}$&	8\\
	&&	NNN&	3&	t$_{2g}$-t$_{2g}$&	12&&\\		

H$_{00}$	&AFM&	NN&	2&	t$_{2g}$-t$_{2g}$&	6	&e$_{g}$-e$_{g}$&	6\\
	&&	NNN&	6	&t$_{2g}$-t$_{2g}$&	20	&t$_{2g}$-e$_{g}$&	10\\

H$_{22}$	&FM&	NN&	1&	t$_{2g}$-t$_{2g}$&	3&	e$_{g}$-e$_{g}$&	3\\
	&&	NNN&	3&	t$_{2g}$-t$_{2g}$&	10&	t$_{2g}$-e$_{g}$&	20\\

\hline
\end{tabular}
\end{table*}

In Table.~\ref{tab:path}, we list the number of these three interactions in other stacking structures. For R$_{22}$-stacking, there are 6 NN t$_{2g}$-t$_{2g}$, 15 NNN t$_{2g}$-t$_{2g}$ and 30 NNN t$_{2g}$-e$_g$ super-super-exchange interactions. Considered the t$_{2g}$-e$_g$ interaction is stronger than t$_{2g}$-t$_{2g}$, the 30 NNN t$_{2g}$-e$_{g}$ interaction should be determinative and the interlayer magnetic coupling of R$_{22}$ should be ferromagnetic. In the R$_{24}$ structure, 12 NN t$_{2g}$-t$_{2g}$, 24 NN t$_{2g}$-e$_{g}$, 7 NNN t$_{2g}$-t$_{2g}$ and 7 NNN e$_{g}$-e$_{g}$ interactions are present. So the interlayer magnetic coupling of R$_{24}$ stacking is decided by the competition between e$_{g}$-e$_{g}$ and t$_{2g}$-e$_{g}$. Considered the larger number of t$_{2g}$-e$_{g}$ interaction but the stronger interaction of  e$_{g}$-e$_{g}$, the R$_{24}$ structure is likely to be weakly ferromagnetic or anti-ferromagnetic, which are consistent with weak ferromagnetism predicted in our work. Recent experiment\cite{Li_NM,science_prob,song_nm} and theory \cite{theory_nl,theory_prb,theory_prm,theory_ssc} indicate the interlayer magnetism of CrI$_3$ in R$_{24}$ order(\emph{C2/m}) is anti-ferromagnetic, which is also consistent with our analysis.

Moreover, the interlayer-distance-dependent magnetic properties of different stacking orders can be also understood by above mechanism. When interlayer distance reduced, interaction between p orbitals of Br atom will become stronger. For example, the interlayer magnetism of R$_{22}$- and H${41}$-stacking are mainly decided by t$_{2g}$-e$_{g}$ due to the weak t$_{2g}$-t$_{2g}$ interaction. The reduced distance will lead the interlayer exchange energy increase monotonously and  ferromagnetism will be strengthened largely. For other stacking structure, in which two comparable exchange mechanisms present, changing layer distance may tune their interlayer magnetic state. The final magnetic order would be determined by the competing between two ferromagnetic and anti-ferromagnetic interactions, like in R$_{20}$- and H$_{22}$-stacking.

It should be pointed out that microscopic mechanisms of interlayer magnetism in bilayer CrI$_3$ have been investigated by several groups\cite{theory_nl,theory_prb,theory_prm,theory_ssc}. However, all these investigations mainly focused on the R$_{22}$ and R$_{24}$ orders of bilayer CrI$_3$ while the interlayer magnetism of other stacking orders including H-type stacking is largely unexplored. In addition, most of the proposed mechanisms \cite{theory_nl,theory_prb} ignore the e$_{g}$-e$_{g}$ interaction, which is important for interlayer magnetism. The present study suggests a competing mechanism based on the principle of symmetry matching and the local orbital direction of Br atoms, which involves not only the t$_{2g}$-t$_{2g}$ and t$_{2g}$-e$_{g}$ interactions but also e$_{g}$-e$_{g}$ interaction. Furthermore, our results can also give a consistent understanding for interlayer magnetism of bilayer CrBr$_3$ with various stacking orders and other vdW bilayer magnets including CrI$_3$.

In summary, we have revealed the underlying mechanism of the stacking order and interlayer magnetism of CrBr$_3$ bilayer. Our results indicate that the stacking energy is closely correlated to the Coulomb potential between the interlayer NN Br-Br atoms. And the stacking-dependent interlayer magnetism can be understood by the competition of super-super-exchange and semi-covalent exchange. The present work provides an insightful explanation for interlayer magnetism of CrBr$_3$, which will be useful to understand the other layered magnetic semiconductors.

\begin{acknowledgement}
This work was supported by National Natural Science Foundation of China (Grant No.11874092 and No.11847157)
the Fok Ying-Tong Education Foundation, China (Grant No. 161005), the Planned Science and Technology Project of Hunan Province (Grant No. 2017RS3034), Hunan Provincial Natural Science Foundation of China (Grant No. 2016JJ2001 and 2019JJ50636), and Scientific Research Fund of Hunan Provincial Education Department (Grant No. 18C0227).
\end{acknowledgement}

\bibliography{ref}
\pagebreak
\clearpage
\begin{suppinfo}
\setcounter{equation}{0}
\setcounter{figure}{0}
\setcounter{table}{0}
\makeatletter
\renewcommand{\theequation}{S\arabic{equation}}
\renewcommand{\thefigure}{S\arabic{figure}}
\renewcommand{\thesection}{S\arabic{section}}
\renewcommand{\bibnumfmt}[1]{[S#1]}
\renewcommand{\citenumfont}[1]{S#1}

The following files are available free of charge.
\begin{itemize}
  \item Fig.~\ref{fig:phonon}: Phonon dispersions of bilayer CrBr$_3$ with R$_{33}$, R$_{15}$ and H$_{41}$ stacking orders.
  \item Fig.~\ref{fig:plusu}: The evolution of interlayer exchange energy with Coulomb interaction parameter U in LDA+U methods.
  \item Fig.~\ref{fig:orb}: The sketch of p orbitals of interlayer NN Br atoms.
\end{itemize}
\begin{figure*}
\centering
\includegraphics[width=0.85\textwidth]{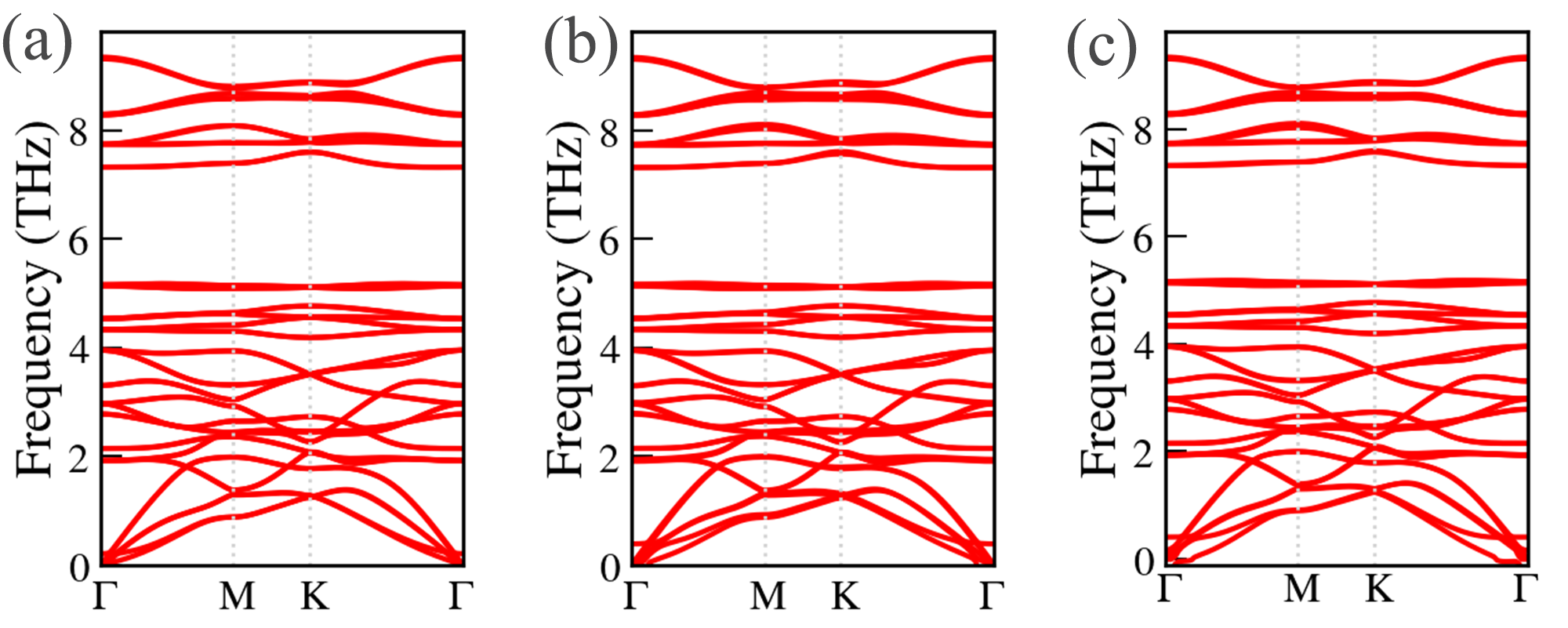}
\caption{\label{fig:phonon}(Color online) Phonon dispersions of bilayer CrBr$_3$ with R$_{33}$, R$_{15}$ and H$_{41}$ stacking orders calculated using the PBE functional.}
\end{figure*}

\begin{figure*}
\centering
\includegraphics[width=0.85\textwidth]{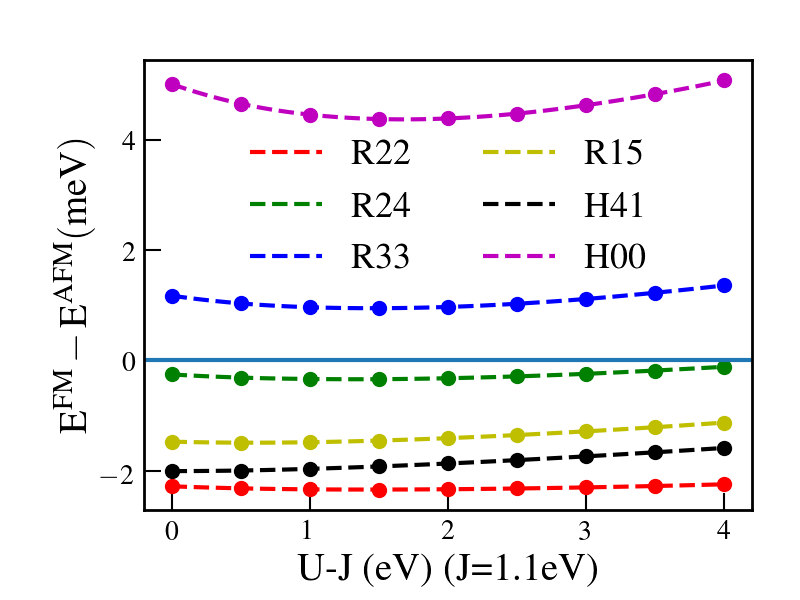}
\caption{\label{fig:plusu}(Color online) The evolution of interlayer exchange energy with Coulomb interaction parameter U for different stacking configurations. }
\end{figure*}

\begin{figure*}
\centering
\includegraphics[width=0.85\textwidth]{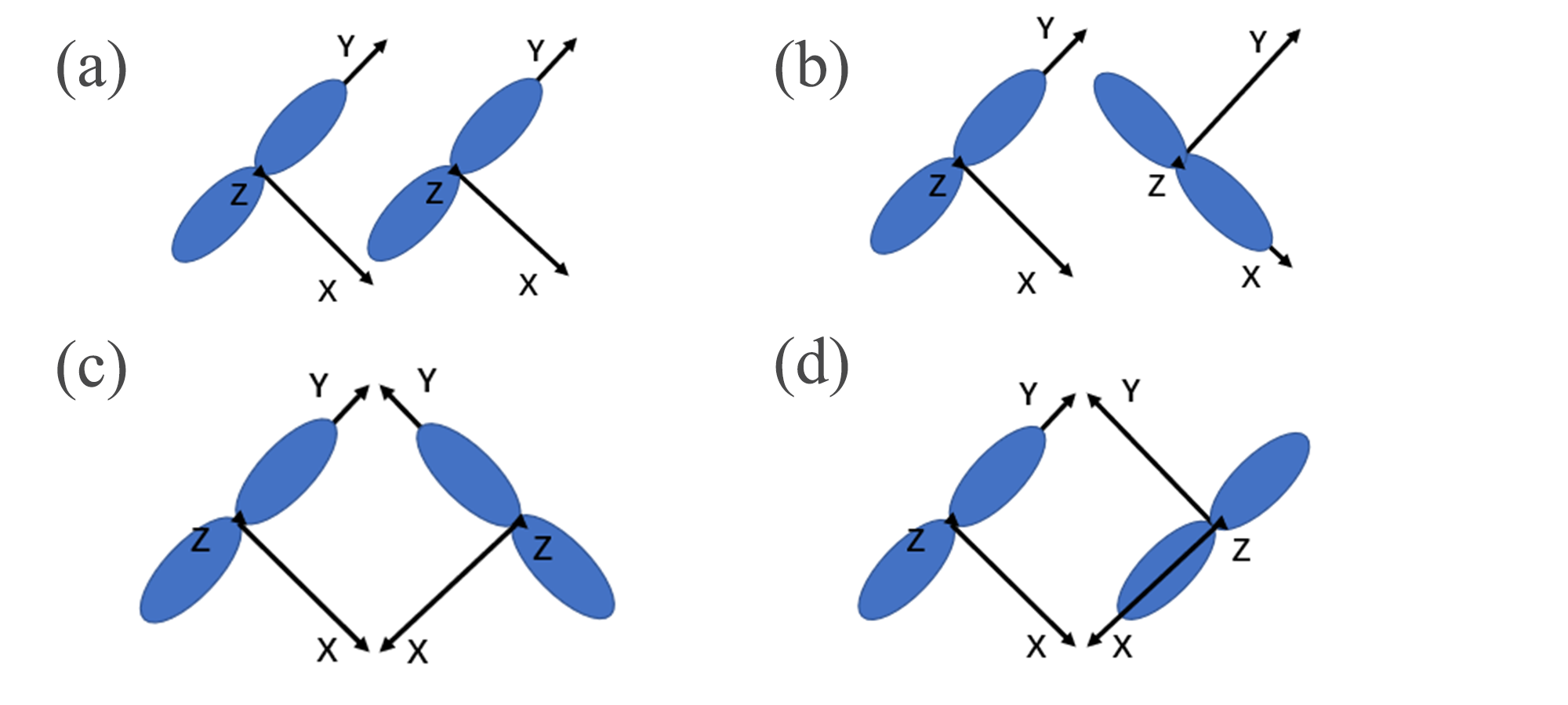}
\caption{\label{fig:orb}(Color online) The sketch of p orbitals of interlayer NN Br atoms. (a) and (b) represent the relative directions of Br atoms in R-type stacking, while the (c) and (d) correspond to that in the H-type stacking. }
\end{figure*}
\end{suppinfo}


\end{document}